\input{aipcheck}

\documentclass[
    ,final            
  ]
  {aipproc}

\layoutstyle{6x9}

\def\aap{A\&A}

\def\apj{ApJ}

\def\mnras{MNRAS}

\def\aapr{A\&ARv}

\begin{document}

\title{Non-thermal processes in colliding-wind massive binaries: the contribution of Simbol-X to a multiwavelength investigation}


\author{Micha\"el De Becker}{
  address={Institut d'Astrophysique et G\'eophysique, Universit\'e de Li\`ege, FNRS, Belgium}
}

\author{Ronny Blomme}{
  address={Royal Observatory of Belgium, Brussels, Belgium}
}

\author{Giusi Micela}{
  address={INAF - Osservatorio Astronomico di Palermo, Palermo, Italy}
}

\author{Julian M. Pittard}{
  address={School of Physics and Astronomy, University of Leeds, UK}
}

\author{Gregor Rauw}{
  address={Institut d'Astrophysique et G\'eophysique, Universit\'e de Li\`ege, Belgium}
}

\author{Gustavo E. Romero}{
  address={Facultad deCiencias Astronom\'{\i}cas y Geofis\'{\i}cas, Universidad Nacional de La Plata, Argentina}
  ,altaddress={Instituto Argentino de Radioastronom\'{\i}a, Buenos Aires, Argentina}
}

\author{Hugues Sana}{
  address={European Southern Obsevatory, Chile}
}

\author{Ian R. Stevens}{
  address={School of Physics and Astronomy, University of Birmingham, UK}
}

\begin{abstract}
Several colliding-wind massive binaries are known to be non-thermal emitters in the radio domain. This constitutes strong evidence for the fact that an efficient particle acceleration process is at work in these objects. The acceleration mechanism is most probably the Diffusive Shock Acceleration (DSA) process in the presence of strong hydrodynamic shocks due to the colliding-winds. In order to investigate the physics of this particle acceleration, we initiated a multiwavelength campaign covering a large part of the electromagnetic spectrum. In this context, the detailed study of the hard X-ray emission from these sources in the SIMBOL-X bandpass constitutes a crucial element in order to probe this still poorly known topic of astrophysics. It should be noted that colliding-wind massive binaries should be considered as very valuable targets for the investigation of particle acceleration in a similar way as supernova remnants, but in a different region of the parameter space.
\end{abstract}

\maketitle


\section{Scientific context}

Stellar winds of massive stars (O-type and Wolf-Rayet) are known to produce thermal (free-free) radio emission. In some cases, synchrotron radio emission has also been identified, revealing the existence of a population of relativistic electrons in such environments \citep{Wh}. Most of these non-thermal radio emitters turn out to be confirmed binaries, or present at least strong indications of binarity \citep[see][for a census of these objects]{debeckerreview}. This discovery provided evidence that an efficient particle acceleration process is at work in colliding-wind massive binaries, and this process is most probably Diffusive Shock Acceleration (DSA) \citep{Pit2,DSACWB}. In the presence of relativistic electrons, a process such as inverse Compton (IC) scattering is expected to be efficient at producing non-thermal X-rays in the form of a power law whose spectral index is directly related to the properties of the population of relativistic electrons. However, the fact that massive stars -- and mostly colliding-wind binaries -- are significant soft thermal X-ray emitters (emission lines from an optically thin plasma), the hard X-ray band needs to be investigated in order to search for this putative non-thermal X-ray emission component \citep{debeckerreview}. The investigation of these non-thermal X-rays is expected to bring crucial information on the physics of particle acceleration in colliding-wind binaries.

\section{The contribution of Simbol-X in this investigation}

\begin{figure}
  \includegraphics[height=.3\textheight]{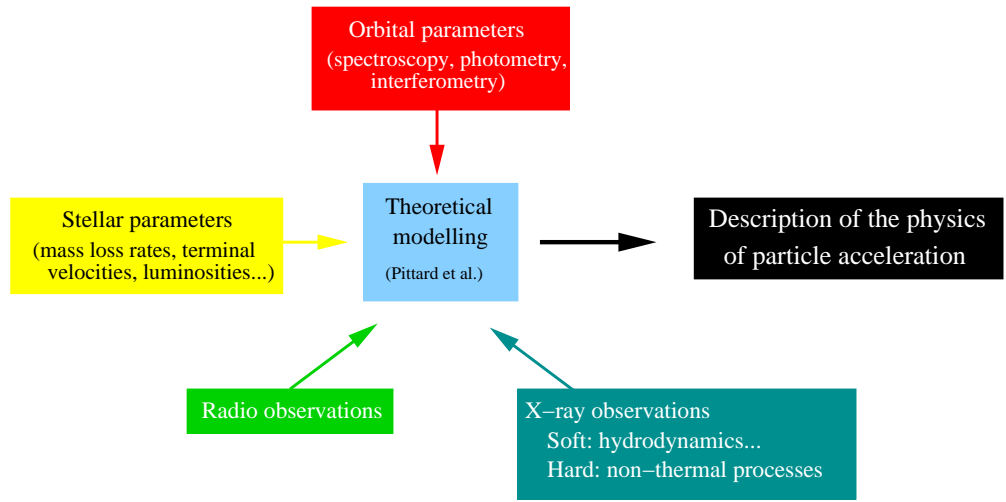}
  \caption{Schematic view of the multiwavelength investigation of the physics of non-thermal processes in colliding-wind massive binaries.\label{multi}}
\end{figure}

The development of recent theoretical models \citep[e.g.][]{Pit2} allows us to probe this phenomenon at least in a few cases where stellar and orbital parameters have been determined through multiwavelength observational studies, as illustrated in Fig.\,\ref{multi}. Stellar parameters can be determined through studies in the visible and infrared domains. In addition, spectroscopic and interferometric studies in the same wavebands are important to determine orbital parameters of colliding-wind binaries. Detailed studies of the soft X-ray emission from these systems further provide crucial information on the hydrodynamics of their colliding-winds. Radio observations provide constraints on the properties of relativistic electrons, but hard X-ray data are strongly needed to lift degeneracies between different sets of parameters \citep[see][]{Pit2}. Even though IC scattering should be at work in the soft X-ray band as well, such an emission component is not expected to be revealed because of the overwhelming thermal emission line spectrum. The non-thermal emission component is indeed expected to be significantly weaker than the thermal one. On the contrary, non-thermal emission should be dominant in the hard X-ray band. Up to now, hard X-ray studies using INTEGRAL failed to detect hard X-rays from non-thermal radio emitters. This lack of detection is most probably attributable to the sensitivity of current hard X-ray observatories. In addition, an angular resolution at the sub-arcmin scale is required to avoid confusion between quite close point sources, such as in the case of the Cyg\,OB2 region \citep{cygint}. For these reasons, Simbol-X is especially appropriate.

\section{Targets for Simbol-X}
\begin{figure}
  \includegraphics[height=.25\textheight]{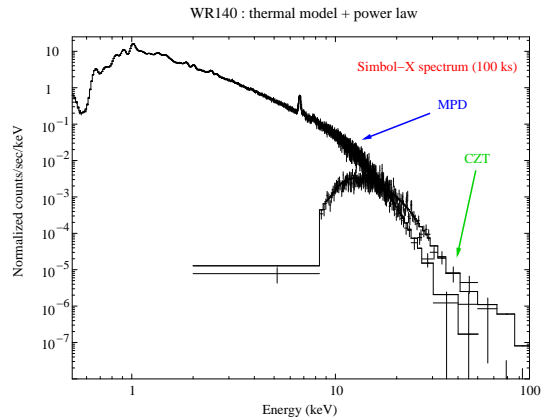}
  \caption{Synthetic Simbol-X spectrum of WR\,140 for an exposure time of 100\,ks.\label{specwr140}}
\end{figure}

According to \citet{debeckerreview}, the catalogue of non-thermal radio emitting massive stars currently includes 16 O-type and 18 WR-type objects. Up to now, none of them has ever been detected in hard X-rays (i.e. above 10\,keV). It is interesting to note that these objects span a wide range of orbital, stellar and wind parameters. Only a few of them have been studied in detail and are likely to benefit from forthcoming modelling efforts in order to investigate the physics of non-thermal processes in massive binaries. Currently, the two best-studied systems are respectively Cyg\,OB2\,\#8A and WR\,140.
\paragraph{Cyg\,OB2\,\#8A} This system consists of an O6I + O5.5III binary, with an orbital period of about 21.9\,d and an eccentricity of 0.24 \citep{Let8a}. A recent detailed investigtion of this system with XMM-Newton revealed a strong soft X-ray emission dominated by the colliding-winds, with significant variations phase-locked with the orbit \citep{DeBcyg8a}. The investigation of a large set of INTEGRAL-ISGRI data led to upper limits for the putative hard X-ray emission from this system \citep{cygint}. With some assumptions on the hard-X-ray flux due to IC scattering, synthetic Simbol-X spectra have been produced, and suggest promising results. For details on these simulations, we refer to \citet{debecker-simbolx}. 
\paragraph{WR\,140} This WC8 + O5-6 binary has a period of about 8\,yr, and is characterized by a large eccentricity (e\,$\approx$\,0.9). The latter is responsible for drastic changes in the physical properties along the orbital cycle, mostly when comparing periastron and apastron. The upper limits on the hard X-ray emission \citep{cygint} allowed to put constraints on the models developed by \citet{Pit2}. On the basis of the latter models, we simulated Simbol-X spectra of WR\,140 for different exposure times and different values for the spectral index of the IC emission component. The soft part of the spectrum was modelled on the basis of a preliminary fit of XMM-Newton spectra of WR\,140 obtained in May 2007 (Obs.ID 055547; PI: M. De Becker). The model used for the global X-ray emission (soft + hard) was a three-temperature model with an additional power law. We selected a photon index of 2.0, along with a normalization parameter corresponding to a hard X-ray flux a factor of 10 lower than the upper limits derived by \citet{cygint}. The synthetic spectrum -- built using the Simbol-X response matrices -- is shown in Fig.\,\ref{specwr140}. The expected quality of the Simbol-X spectrum is sufficient to model in detail the non-thermal X-ray emission component from this system.

\section{Conclusions}

We can formulate the following conclusions:
\begin{enumerate}
\item[1.] Colliding-wind binaries can be considered as valuable targets for future observations with Simbol-X.
\item[2.] Our aim is not just to detect a putative hard X-ray emission component from such systems. We intend to determine fluxes and photon indices in order to study the properties of the relativistic electrons, therefore probing the physics of particle acceleration is these environments. In addition, as we are dealing sometimes with quite eccentric systems, we consider that it would be more telling to study these systems at various orbital phases in order to investigate phase-locked variations of the non-thermal emission. We emphasize however that the physics we intend to study using Simbol-X is related to relativitic electrons (i.e. leptonic processes). The physics of relativistic protons (i.e. hadronic processes) would require to investigate much higher energies such as those reachable for instance by the Fermi telescope.
\item[3.] Targets are already identified, including a few for which many crucial parameters have been identified notably in the context of our multiwavelength investigation of non-thermal processes from massive binaries.
\end{enumerate}


\begin{theacknowledgments}
MD acknowledges the Fonds National de la Recherche Scientifique (FNRS) for financial support, along with the organizing committee of this conference for giving the opportunity to present this science topic.
\end{theacknowledgments}

\bibliographystyle{aipproc}   


\end{document}